# A note on semidiurnal motions in the deep Mediterranean

**by Hans van Haren**

Royal Netherlands Institute for Sea Research (NIOZ), P.O. Box 59, 1790 AB Den Burg, the Netherlands.
e-mail: hans.van.haren@nioz.nl

**Abstract**

The Mediterranean Sea is known for its limited tidal motions. For example, surface barotropic tidal elevations have an amplitude of about 0.1 m in the Northwestern Mediterranean. Nevertheless, small significant peaks occur at semidiurnal frequencies in spectra of temperature variance measured near the 2450-m deep seafloor at a multiple-line mooring. These semidiurnal signals appear to be surface tides commonly measured by pressure sensors, but are only observable in temperature under near-homogeneous conditions when buoyancy frequency N < f, the inertial frequency. After transfer of pressure to temperature units via the local adiabatic lapse rate, the observed internal-wave temperature signals may thus be corrected for $1.5\times10^{-5}$-°C amplitude semidiurnal, predominantly lunar barotropic tides. The remaining baroclinic tides are embedded in the broad and featureless inertio-gravity wave 'IGW' band, with small but significant energy enhancement near its boundaries including the precise semidiurnal frequency, also under tenfold-larger energetic stratified-water conditions. The similar IGW bands under both stratification conditions suggest slanted convection during stratified-water conditions, as the stratification is near-homogenous in the direction slanted to the vertical. The use of multiple instruments improves spectral statistics at frequencies exceeding IGW, and including the IGW band under some near-homogeneous conditions but not under stratified-water conditions during which more coherent motions prevail.

## 1 Introduction

The deep Western Mediterranean Sea may be void of sunlight and its waterflow may be relatively slow at total speeds of about 0.05 m $s^{-1}$ (van Haren et al. 2026), yet it is not stagnant but full of variable flows (e.g,, Testor and Gascard 2006). For example, above the generally flat seafloor in the vicinity of its northern continental shelf, dynamical processes include a large-scale boundary flow (e.g., Crepon et al. 1982), with meanders and eddies varying at 10−30 day mesoscales and 1−10 day sub-mesoscales (Albérola et al. 1995).

Shorter-period deep-sea flow variations involve internal waves, notably at near-inertial scales, as transients following passages of atmospheric disturbances induced by, e.g., varying winds from the nearby mountain ranges like the Alps. Near-inertial motions can penetrate as propagating ‘internal waves’ into the deep sea, also via trapping by (sub-)mesoscale eddies (Kunze, 1985). They may set-up shorter scale internal waves that eventually dissipate their energy through irreversible turbulence when they break. In the open ocean, tides at predominantly semidiurnal frequencies like lunar $M_2$ and solar $S_2$ $\approx 2\Omega$ are an important generator of internal waves.

In contrast with most seas and oceans however, tides are generally weak in the Mediterranean, with a few local exceptions. The weak tides in most of the Mediterranean result from poor resonance conditions so that they reflect direct generation by the Moon-Sun system. In the Northwestern Mediterranean typical sealevel amplitudes are 0.1 m (https://www.tide-forecast.com/locations/Toulon-France/tides/latest). These ‘barotropic’ surface tides, which have unattenuated amplitude from surface to bottom, may generate internal ‘baroclinic’ tides, which have distinct amplitude and phase variations over short distances, when vertical density-stratification and seafloor-topography conditions are favourable.

Although the weak tides reduce the amount of internal-wave energy by about 60% (Wunsch and Ferrari 2004), internal-wave breaking constitutes a non-negligible generator of turbulence besides geothermal heating in the deep Mediterranean (e.g., van Haren et al. 2014; Ferron et al. 2017; van Haren et al. 2026). This internal-wave breaking may involve motions at semidiurnal frequencies, which however are not associated with deterministic tides but part of a broader range of internal wave motions under very weakly stratified conditions.

In the open ocean, the local mean buoyancy frequency reflecting stable vertical density stratification has values of $N >> f$, where $f = 2\Omega\sin\varphi$ denotes the inertial frequency or vertical Coriolis parameter at latitude $\varphi$ and $\Omega$ the Earth rotational frequency. Under such conditions, the frequency ($\omega$) range of freely propagating linear internal waves has bounds $\omega \in [f, N]$. In the deep Mediterranean however, stratification is very weak $N = O(f)$ compared to that in most of the deep ocean. Under such weakly stratified conditions (van Haren and Millot 2003), the range $[f, N]$ of freely propagating internal waves spreads from semidiurnal frequencies near $2\Omega$ well into the sub-inertial, $\omega < f$, sub-mesoscale range as not only gravity but also Earth's rotational momentum plays a role as restoring force (LeBlond and Mysak 1978): inertio-gravity waves 'IGW'.

This is because the common internal-wave band $f < \omega < N$, for strong stratification with $N >> f$ (LeBlond and Mysak 1978), becomes modified by the effect of traditionally neglected horizontal Coriolis parameter $f_h = 2\Omega\cos\varphi$ under weakly stratified conditions. For $N = O(f)$, minimum IGW-bound $\omega_{min} \leq f$ and maximum IGW-bound $\omega_{max} \geq 2\Omega$, if $N < 2\Omega$, are functions of N, $\varphi$, and of the direction of wave propagation, so that (LeBlond and Mysak 1978; Gerkema et al. 2008),

$$\omega_{max}, \omega_{min} = (A \pm (A^2 - B^2)^{1/2})^{1/2}/\sqrt{2}, \quad (1)$$

in which $A = N^2 + f^2 + f_s^2$, $B = 2fN$, and $f_s = f_h\sin\alpha$, $\alpha$ the angle to zonal direction. One effect of $f_h$ is turbulent-convection in slantwise direction in the vertical-meridional z,y-plane, so that apparently stable stratification observed in z-direction may actually reflect homogeneous or unstable conditions in the tilted plane of planetary vorticity, except at the north pole (e.g., Straneo et al. 2002; Sheremet 2004). Another effect of $f_h$ is that semidiurnal tidal frequencies are always included in IGW, albeit for meridional propagation direction only.

Internal-wave bounds may also vary due to (sub-)mesoscale motions. Local time- and space-varying horizontal waterflow differences such as in meanders and eddies can generate the vertical component of relative vorticity $\zeta = \partial v/\partial x - \partial u/\partial y$ with reported amplitudes of up to $|\zeta| = f/2$ around mid-depth in the Western Mediterranean (Testor and Gascard 2006). This addition to planetary vorticity f is suggested to widen the 'effective' near-inertial band (Kunze 1985).

In this note, the contribution of barotropic and baroclinic tides to the potentially widened IGW in the deep Mediterranean is investigated using observations made at a large three-dimensional mooring array mainly holding many temperature 'T-'sensors at multiple vertical lines for improved statistics. Pressure information is used to separate dynamically unimportant barotropic tides, which appear as O($10^{-5}$ °C) amplitudes, from temperature records. This detailed study follows a general overview of results from this array presented in van Haren et al. (2026).

## 2 Materials and Methods

Nearly half-a-million cubic meters of deep Mediterranean seawater was measured every 2 s using 2925 self-contained high-resolution NIOZ4 T-sensors, which can also record tilt and compass. A scheme of the experiment is given in Appendix A, details of mooring deployment can be found in van Haren et al. (2021). The mooring site was about 40 km south of Toulon (France), where $f = 9.9\times10^{-5}$ $s^{-1}$ = 1.36 cpd (cycles per day). Temperature-only sensors were taped at 2-m intervals to 45 vertical lines 125-m tall. In addition, two tilt-temperature sensors were attached near the top and bottom of each line, which was tensioned to 1.3 kN by a single buoy above.

Three buoys, distributed over about 50 m over the mooring-array, held a downlooking acoustic single-point Nortek AquaDopp current meter that recorded waterflow and, important for establishing barotropic tidal surface elevations, pressure at a rate of once per 600 s. Although the amplitudes of deep-sea waterflow are seldom larger than 0.05 m $s^{-1}$ (van Haren et al. 2026), instrumental noise occurs at $\omega$ > 4 cpd, so that lower-frequency tidal, inertial, sub-mesoscale and mesoscale motions are well resolved. The waterflow was thus sampled at the height of the uppermost T-sensor at h = 126 m above seafloor.

Probably due to a format error, the T-sensors switched off unintentionally after maximum 20 months of data-recording. Tilt-temperature sensors recorded only 5.5 months of data, which data are not considered here. As with previous NIOZ4 T-sensors (for details see van Haren 2018), the T-sensors' individual clocks were synchronized to a single standard clock every 4 hours, so that all T-sensors were recording data within 0.02 s. About 25 T-sensors failed mechanically. During post-processing, some 20 extra T-sensors are not further considered due to general electronics (noise) problems. Data from these sensors are not considered in spectral analyses and are linearly interpolated between data from neighbouring sensors in other analyses. One near-bottom T-sensor failed. Instrumental bias was removed via vertical smoothing (van Haren et al. 2005; van Haren 2018) and via low-pass filtering with cut-off at 0.05–0.1 cpm (cycles per meter) depending on the stratification rate. In addition, because vertical temperature (density) gradients are so small in the deep Mediterranean, reference was made to periods of typically one hour duration that were homogeneous with temperature variations less than instrumental noise level of about $3\times10^{-5}$ °C (van Haren 2022). Such quasi-homogeneous periods existed on days 350, in 2020, 453 and 657, both −366 in 2021, in the records. The reference was necessary for inter-calibration and bias-removal of the sensors. Despite the low precision of $<10^{-4}$ °C, absolute accuracy of $<10^{-3}$ °C is not achievable with these sensors over 600 days of observations.

For reference, a single shipborne Conductivity-Temperature-Depth 'CTD' profile was obtained within 1 km from the site of large-ring mooring, during the deployment cruise.

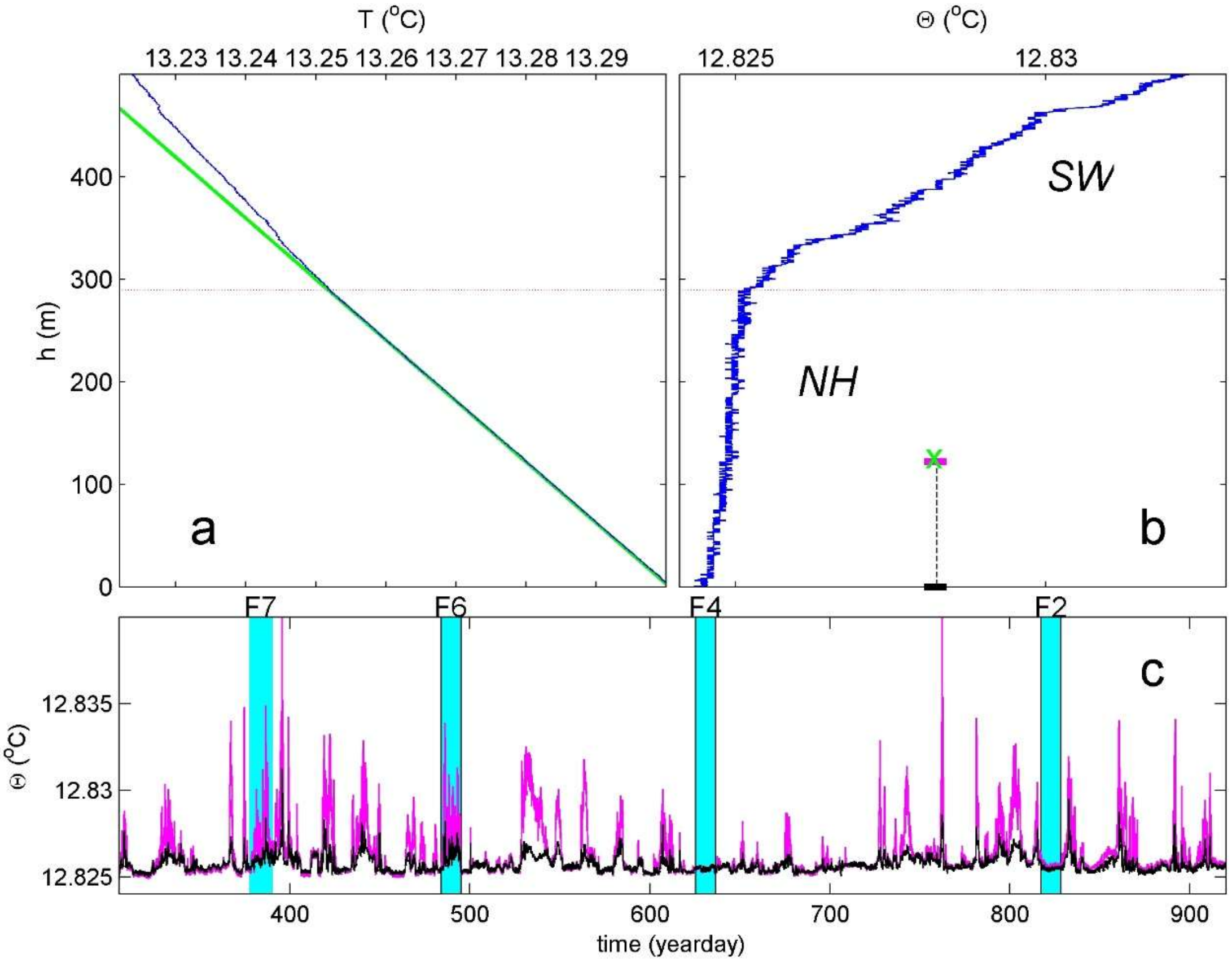

**Figure 1.** Data overview from deep Mediterranean waters at the site of the large-ring mooring. (a) In situ temperature from lower 500 m of shipborne CTD profile. In green, the local adiabatic lapse rate is given starting at the temperature observed at h = 0.5 m above seafloor. (b) Pressure-corrected

Conservative Temperature (IOC et al. 2010), which has the same dynamical properties as potential temperature, is presented together with locations of uppermost (magenta) and lowest (black) moored temperature ‘T’-sensors, and (green x) moored current meter ‘CM’. SW = stratified-water, NH = near-homogenous conditions. (c) Entire 600-d time series sub-sampled at once per 6 s of uppermost and lowest moored T-sensor data from one line, with episodes in Figs 2, 4, 6, and 7 highlighted. Time is given in days of 2020 (+366 in 2021).

## 3 Results

The lower 500 m of the CTD-profile shows measurable weak stratified-water ‘SW’ conditions down to $h \approx 300$ m from the seafloor, and near-homogeneous ‘NH’ conditions closer to the seafloor (Fig. 1a, b). The transition between the two conditions is quite abrupt in vertical density stratification. However, in both cases stratification is rather weak, with 100-m-scale buoyancy frequency $N \approx 2f$ under SW and $N < 0.5f$ under NH. The generally weak stratification implies that the adiabatic lapse rate $\Gamma$ of compressibility dominates the variation in temperature with depth, especially for $h < 300$ m (Fig. 1a). $\Gamma$ is a function of local salinity, temperature and pressure p, and amounts,

$$\Gamma = d\Theta/dp = 1.68 \pm 0.01 \times 10^{-8}\ °C\ m^2\ N^{-1}, \quad (2)$$

for the range in Fig. 1a. After transfer of pressure to temperature units via the local adiabatic lapse rate, observed internal-wave temperature signals may thus be corrected, via ‘pressure-data filtering’, for $1.5\times10^{-5}$-°C amplitude semidiurnal barotropic tides (of 0.09-m amplitude surface elevations for Toulon harbour).

Weak stratification with $N = O(f)$ is typical for 2000–2500-m deep waters in the Mediterranean, but is generally found at much greater depths around 5000 m in oceans. This may have to do with the lack of tides in the Mediterranean, and thereby reduced internal wave energy (by about a factor of 2 compared to the ocean; Wunsch and Ferrari 2004) with associated reduced turbulent mixing via internal-wave breaking. Also, the Mediterranean Sea is an almost closed basin, nearly without connections to other basins whereby wave propagation from far away sources is likely reduced, in contrast with oceans.

The Mediterranean CTD-profile is representative for local conditions within the vertical range $h < 126$ m of moored instrumentation (cf. Fig. 1b) during about half of their time underwater. During the remainder, SW conditions reach the moored instrumentation, either from above or from the side. They provide 124-m vertical temperature differences of up to about 0.01°C instead of $<2\times10^{-4}$ °C under NH conditions (Fig. 1c). The conditions have a typical timescale of 10–20 days, and vary rather abruptly from one to the other. This variability is attributed to (sub-)mesoscale eddy activity related with variations in continental-boundary flow by atmospheric forcing. Switches between the two deep-sea conditions are found year-around at the mooring site, with some increase in activity during winter (van Haren et al. 2026). Largest turbulence values are estimated under SW conditions, when relatively warm waters are pushed in slantwise from above, either via eddies and/or IGW.

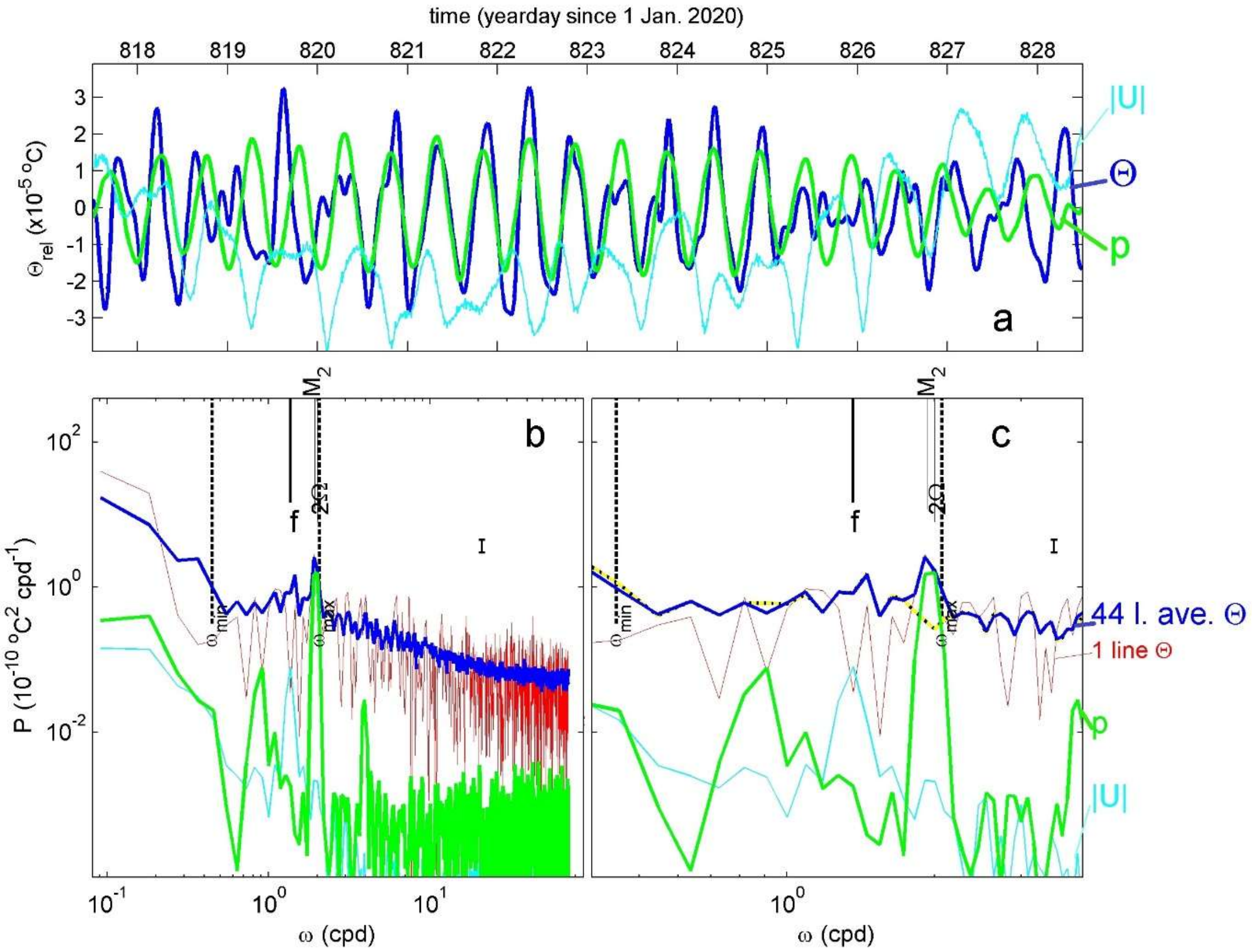

**Figure 2.** Magnification of 11 days of data under near-homogeneous 'NH' conditions dominated by geothermal heating observed at near-bottom h≈2-m T-sensors and upper-range h=126-m CM's, all (sub-)sampled at once per 600 s. (a) Time series comparison between records relative to their mean value of semidiurnal band-pass filtered 'bpf' temperature (average over 44 lines; one T-sensor was not working properly), bpf pressure (transferred to temperature via the local adiabatic lapse rate; average over 3 CM's), and low-pass filtered waterflow amplitude (arbitrary units; average over 3 CM's). (b) Energy spectra for records in a., together with temperature from one line. The error bar is for 44-line smoothed data. Besides inertial frequency f, semidiurnal 2Ω, and lunar $M_2$, inertio-gravity wave 'IGW' bounds [$\omega_{min}$ $\omega_{max}$] are indicated for buoyancy frequency N = 0.5f. (c) One-decade zoom on IGW from b., with the addition of pressure-data filtered 44-line mean temperature spectrum (dashed).

In order to investigate spectral characteristics with reduced statistical uncertainty one could average over instruments of all mooring lines and over the entire period of 600 days. However, such averaging, spectral smoothing, would blur the temperature variance and obscure lesser variance under NH conditions by that under SW. Alternatively one could select and average values only under NH conditions, and likewise under SW conditions, but the lengths and characteristics vary per NH period and per SW period. Therefore, the detailed analysis below is done for several 11-day periods, a common maximum length under NH- and SW conditions, which are considered representative for different conditions. The spectral smoothing is then obtained by averaging over the multiple mooring lines.

Over a relatively long 11-day episode under NH-conditions that appears to be dominated by geothermal heating from below, semidiurnal variations in pressure, which reflect surface barotropic tidal elevations, match in size near-bottom temperature variations after considering (2), see Fig. 2. The temperature-variation amplitude is $\Theta \approx 2\times10^{-5}$ °C, while that of pressure $p \approx 10^3$ N m$^{-2}$ and of waterflow velocity $|U| \approx 1.5\times10^{-2}$ °C m s$^{-1}$. The focus is on temperature from T-sensors closest to the seafloor, because of all T-sensors these are least collecting (sub-)mesoscale, baroclinic internal wave and turbulence motions. Lesser diurnal and fourth-diurnal peaks in pressure do not stand out from the broad variance in temperature. Diurnal peaks in year-averaged pressure spectra (not shown) reflect varying dominant frequencies at tidal frequencies $O_1$, $K_1$, with a shoulder at $S_1$ possibly reflecting a tidal signal mixed with sea-breeze effects at the mooring site about 20 km south of the nearest island. In contrast, waterflow variations show less semidiurnal variation with time, and have dominant responses around f and at sub-mesoscale frequencies instead. At the inertial frequency, pressure variations are very weak, indistinguishable from noise level, and temperature variations are part of a broadband, above-noise continuum.

The 44-line averaged temperature spectra show a considerable smoothing with reduced statistical uncertainty over the 1-line spectrum at most frequencies, except at the semidiurnal tidal frequency (Fig. 2b, c). This exception suggests a deterministic signal rather than a quasi-randomly distributed spectral content. Band-pass filtered temperature matches ($\Gamma$-transferred) pressure semidiurnal-lunar $M_2$ peaks thus well (Fig. 2a), especially in the mid-half of the time series, that the barotropic surface tide explains about 75% of deep-sea semidiurnal temperature variance in this narrow frequency band. The barotropic tide can be relatively easily removed, from this record: The pressure-data filtered temperature spectrum in Fig. 2 lacks a peak around $M_2$.

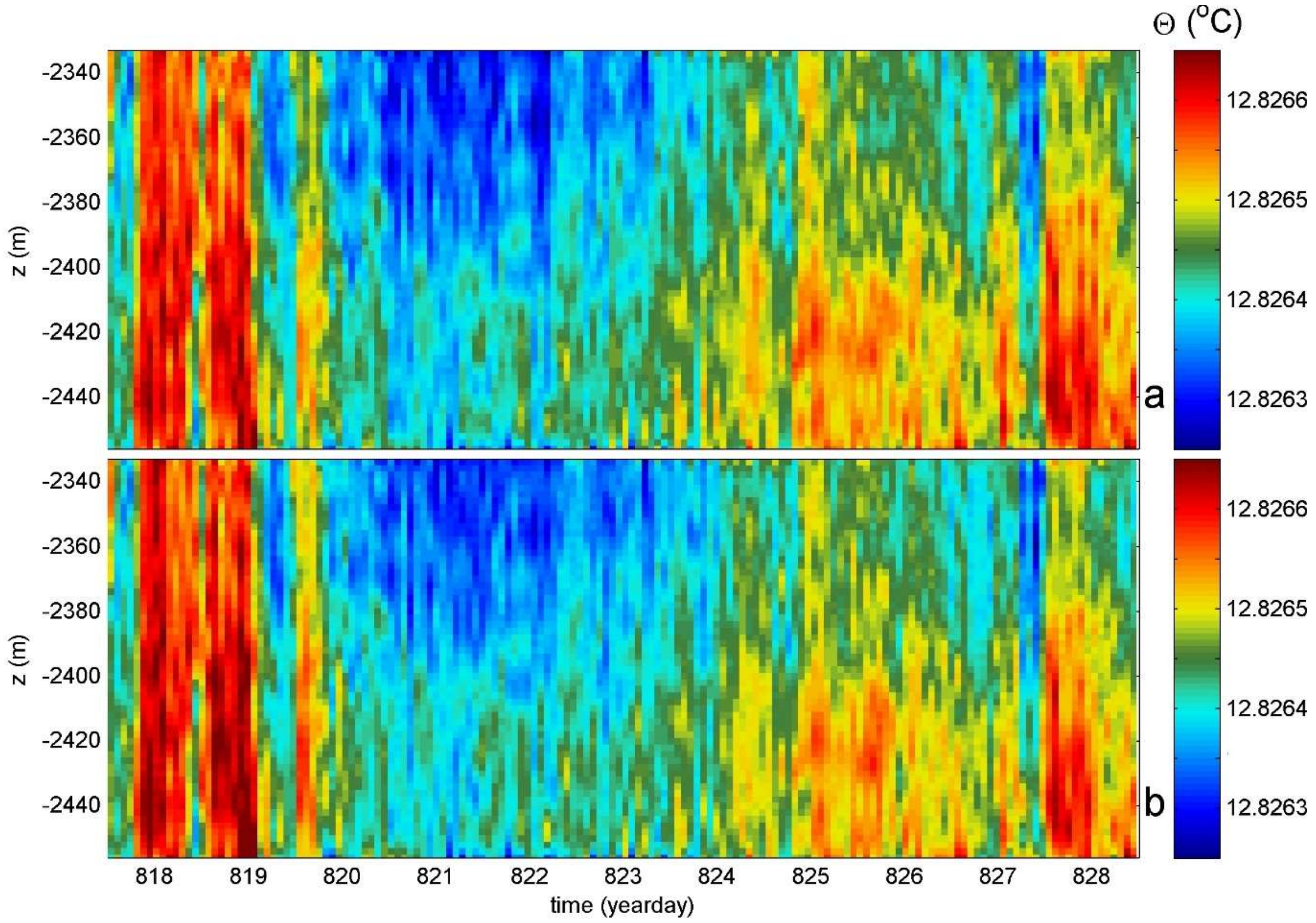

**Figure 3.** Effect of pressure-data filtering on 600-s sub-sampled temperature data from one line for NH in Fig. 2. (a) Time-depth plot sub-sampled at once per 6000 s of Conservative Temperature after common post-processing (van Haren 2018). (b) Data from a., after additional barotropic-tidal filtering using pressure data.

However, in time-depth temperature images the removal of the deterministic-narrowband barotropic tide, which is the same for all vertical levels, does not very clearly show, except in the center of the record (Fig. 3). This is because the dominant barotropic $M_2$-motions are embedded in broadband, less deterministic and more stochastic, baroclinic internal wave motions, which are not necessarily the same at vertical levels, and which fill the near-bottom IGW-band of temperature variance almost like flat white noise (Fig. 2b, c). As the example is dominated by turbulent convection from below governed by geothermal heating, apparent near-tidal columns are warmest near the seafloor (Fig. 3). The removal of the barotropic tide smooths the edges of the convection plumes, which have a dominant frequency varying between f and about $2\Omega \approx \omega_{max}$.

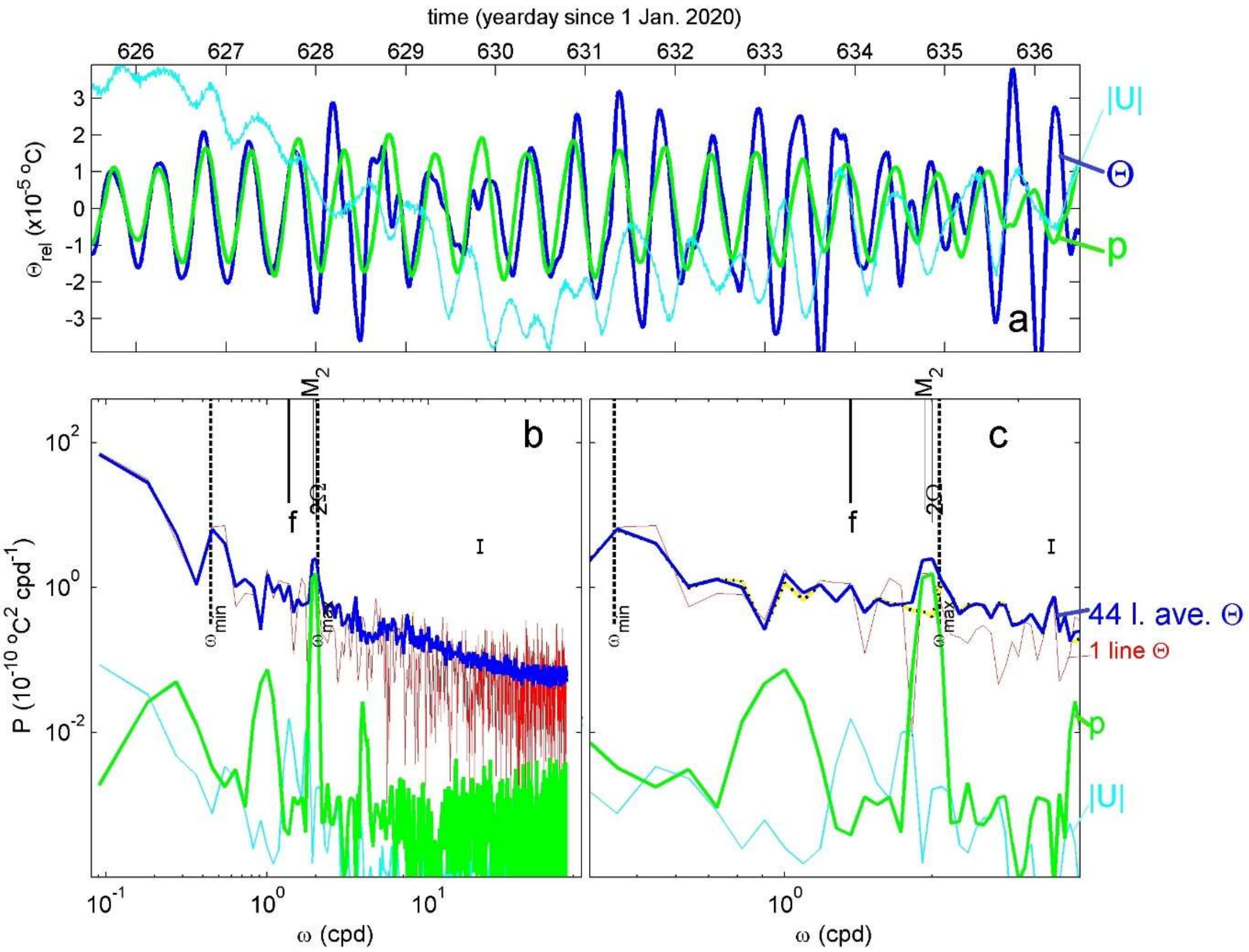

**Figure 4.** As Fig. 2, but for different NH-conditions dominated by h > 125-m vertical columns.

From a different 11-day example under NH-conditions that appears to be dominated by >125-m tall vertical "columns", the IGW bounds contain most IGW-temperature variance in the otherwise weakly sloping band of baroclinic waves (Fig. 4). Compared with Fig. 2, near-inertial waterflow motions are slightly reduced in this example, or spread over a wider band of (super-inertial) frequencies. As in the previous example, barotropic signals take up a considerable part, explaining about 50% of the variance of semidiurnal near-bottom temperature signals. Their removal via the pressure record and (2) has slightly more visual effect in the time-depth plot of Fig. 5, compared to that of Fig. 3, not only in the center of the time series, but also near the beginning. In contrast with Fig. 2, the entire IGW- and sub-inertial bands show less smoothing for the 44-line average spectrum compared with the 1-line spectrum (Fig. 4b, c). This suggests more coherent motions at these frequencies than in the example of geothermally dominated turbulence convection, possibly associated with the somewhat larger temperature variance in the IGW band of Fig. 4.

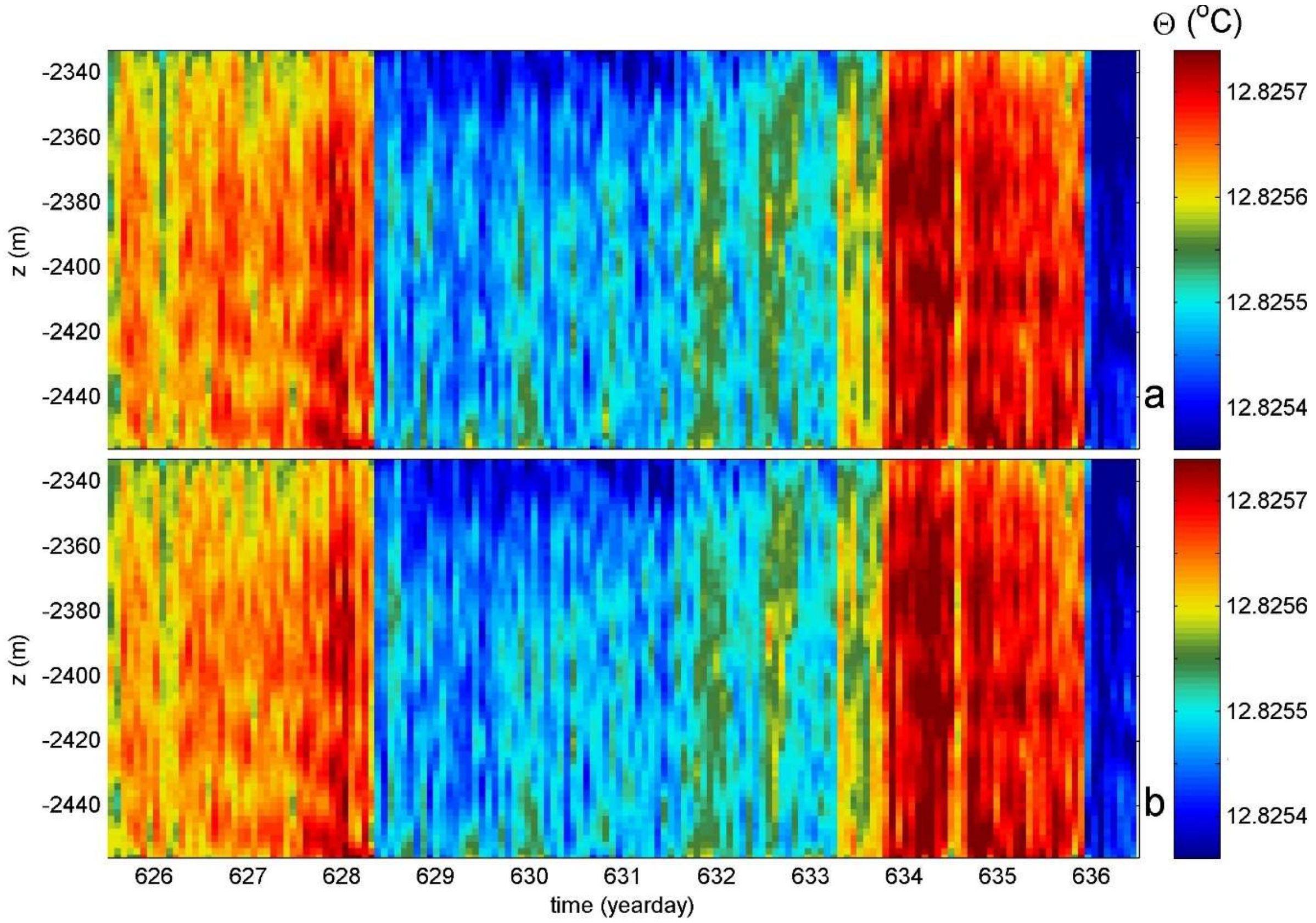

**Figure 5.** As Fig. 3, but for NH of Fig. 4.

The removal of barotropic signals via pressure record is unnecessary under SW conditions, when temperature variations from the IGW- and sub-mesoscale bands are larger by one order of magnitude, two orders in variance, compared with NH conditions, see the 11-day example under SW conditions in Fig. 6.

The IGW band is almost flat in variance distribution (Fig. 6b, c), like under NH conditions in Fig. 2. Compared with the 1-line spectrum, the 44-line average spectrum is only smoothed for ω > 10 cpd, which demonstrates a considerable extent in frequency range of coherent motions, in comparison with Figs 2 and 4. With the increased overall IGW variance in temperature, the waterflow spectrum shows a smaller inertial peak than in Fig. 4 besides increased sub-mesoscale/sub-inertial-IGW activity.

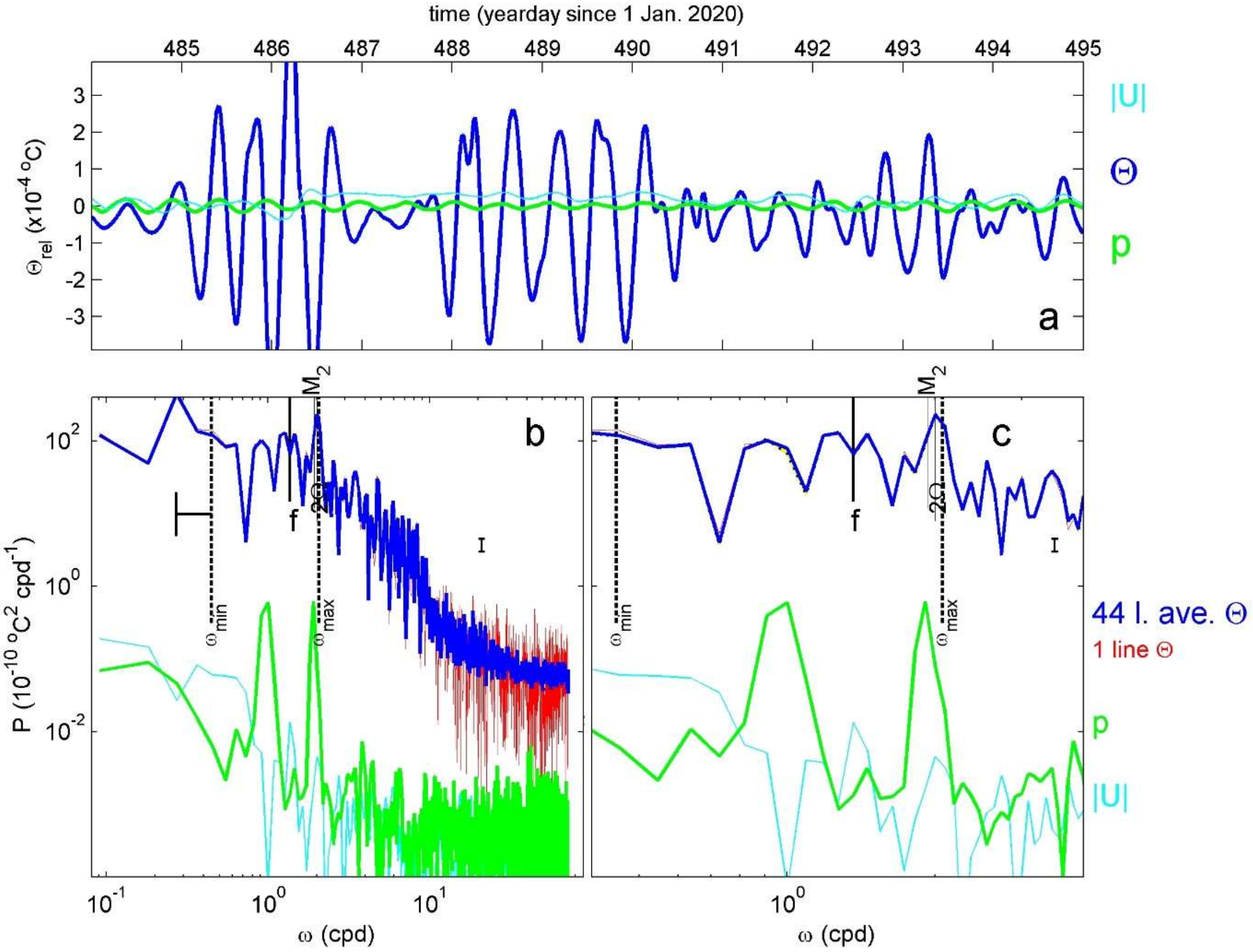

**Figure 6.** As Fig. 2, but for stratified-water conditions. IGW-bounds are for N = 0.5f as under NH, not for N = 2f as for vertical stratification under SW. The 90°-rotated-T extension to the $\omega_{min}$-bound in b. is for N = 0.3f (see text). The y-axis range in a. is 10 times larger than in Fig. 2a.

Semidiurnal pressure and temperature signals not only differ strongly in amplitude, but also in phase (Fig. 6a). As a result, temperature is dominated by baroclinic internal wave signals, which, however, still show a small statistically-significant semidiurnal peak (Fig. 6b,c). Essentially, this small peak is not at $M_2$, but around $2\Omega \approx \omega_{max}$ (for N < 0.5f). The small temperature peak near the upper IGW bound is surprising considering that mean vertical N ≈ 2.2f ≈ 3 cpd, a clear shift to higher frequencies in comparison with the NH examples in Figs 2 and 4. However, the relative peak at about 0.27 cpd (Fig. 6b) may be equivalent to $\omega_{min}$ for N = 0.3f (for which $\omega_{max}$ = 2.03 cpd). This difficult-to-measure small N suggests that the apparently stable SW-conditions are actually very weakly stratified, in their direction of turbulent convection that is seemingly slanted away from gravity.

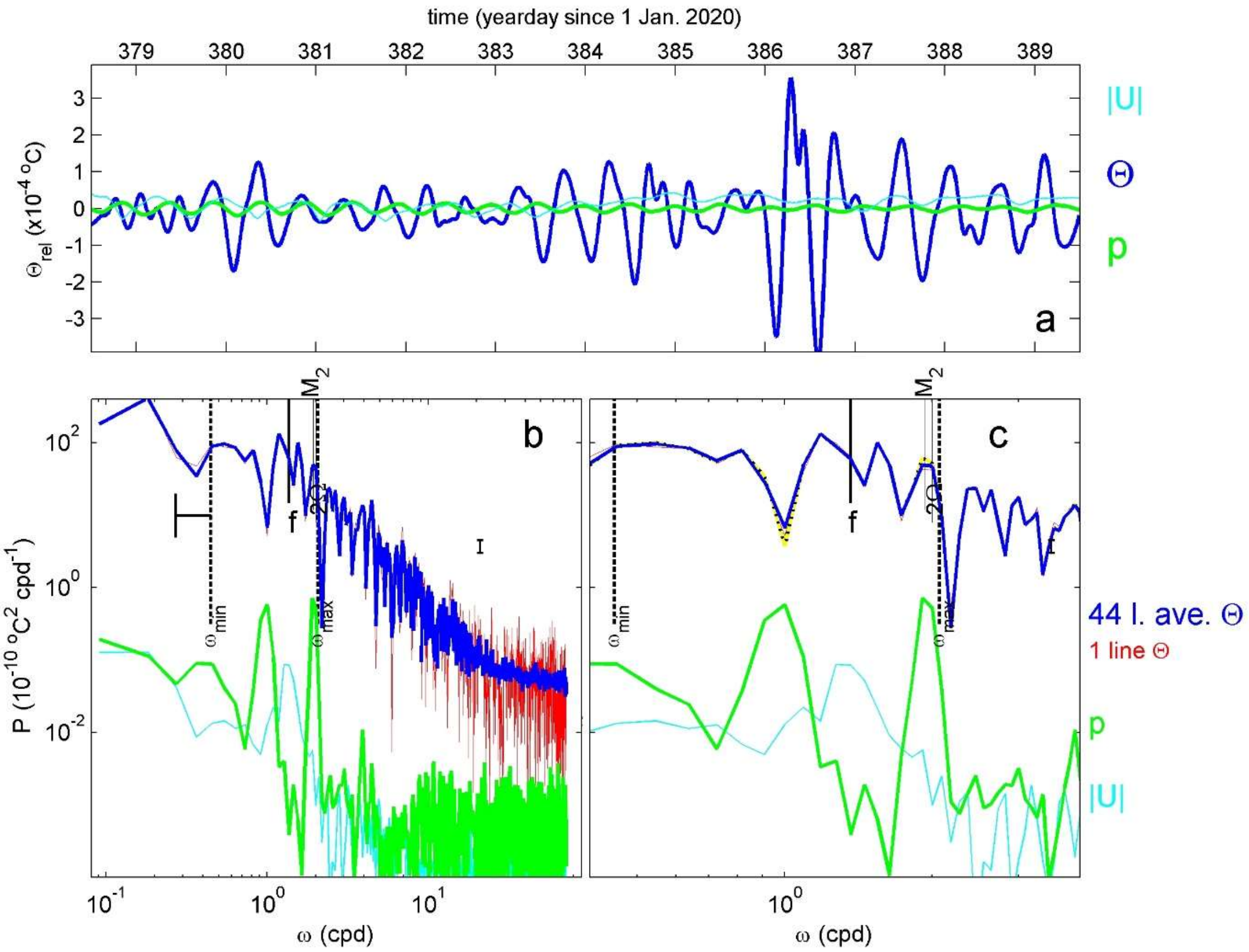

**Figure 7.** As Fig. 6, but for a different episode under SW conditions, with larger near-inertial peak in waterflow spectrum like in Fig. 2.

A different episode under SW conditions demonstrates similar characteristics (Fig. 7). Like in Fig. 6, the temperature variance is large in the IGW band but almost horizontally flat distributed, over a wider IGW range than expected from the overall large-scale vertical N. Small differences with Fig. 6 exist, like a shift of a sub-mesoscale peak towards even smaller frequencies of about 0.2 cpd, a less pronounced semidiurnal peak and slightly enhanced variance around f. The waterflow shows a more distinct peak at f. This example largely confirms the observation in Fig. 6 leading to a hypothesized potential dominance of slantwise convection, as has been observed in short movies (van Haren et al. 2026).

Slantwise turbulent convection under SW conditions may lead to a widening of the IGW range compared to that inferred from vertical stratification, because stratification is close to homogeneous in the slanted direction. Following the present deep Mediterranean observations, apparently the IGW-widening occurs with some enhancements near its bounds including baroclinic semidiurnal internal wave motions. As the convection is governed by highly nonlinear processes, the associated relatively strong turbulence that is elevated by one order of magnitude over deep open-ocean values (van Haren et al. 2026) seems to be important for the replenishment of nutrients in the deep Mediterranean. The sub-inertial extent of the IGW-band may include sub-mesoscale eddies, of which the precise interaction with IGW is unknown.

**4 Discussion and Conclusions**

In future research utilizing a full year of data from the deep Northwestern Mediterranean, there is no need to be much concerned that barotropic surface tidal signals spoil baroclinic internal wave signals, e.g. in scalar observables like from moored temperature sensors. The barotropic tides are weak, and only show semidiurnal variations O($10^{-5}$ °C) in temperature records under near-homogenous conditions. Such small temperature signals are successfully corrected using pressure information and local adiabatic lapse rate.

Indirectly though, barotropic tides potentially may have effect after sufficient energy transfer to baroclinic IGW. In the present data such a transfer is not well observed, as under near-homogeneous conditions semidiurnal temperature variations are mainly attributable to barotropic sealevel variations. Under stratified-water conditions, no dominant lunar semidiurnal signals are observed in temperature. Instead, the observed semidiurnal variations are associated with the maximum inertio-gravity wave bound. Likewise, no semidiurnal lunar peak is observed in waterflow spectra.

Under NH conditions, the observed broadband spread of temperature variance throughout IGW is attributable to $N < f$, confirming non-traditional internal-wave frequency bounds. Small peaks occur at the bounds and reflect some wave trapping. The lower IGW bound extends well into the sub-mesoscale range. No effect of eddies in terms of relative vorticity is observed, although waterflow demonstrates a small peak around 1.2f.

Under SW conditions, the elevated temperature variance shows basically the same IGW spread as under NH conditions including enhancements near IGW bounds, which does not comply with local mean vertical N. Instead, it reflects an IGW for very weak $N \approx 0.3f$ presumed in a slantwise direction.

Or, it reflects a band-widening by local relative vorticity as the waterflow spectrum suggests energy enhancements around 0.5f and 1.5f, albeit without strong evidence during different episodes.

**Acknowledgments** Captains and crews of R/V Pelagia are thanked for the very pleasant cooperation. NIOZ colleagues notably from the NMF department are especially thanked for their indispensable contributions during the long but very pleasant preparatory and construction phases of the large-ring mooring to make this unique sea-operation successful. I am indebted to colleagues in the KM3NeT Collaboration, who demonstrated the feasibility of deployment of large three-dimensional deep-sea research infrastructures. V. Bertin helped with local organization in La Seyne-sur-mer.

**Author contributions** HvH performed all analyses and wrote the manuscript.

**Funding** This research was supported in part by NWO, the Netherlands Organization for the advancement of science.

**Data availability** Current meter and CTD data are available from van Haren (2025): “Large-ring mooring current meter and CTD data”, Mendeley Data, V1, https://doi.org/10.17632/f8kfwcvtdn.1. Processed temperature data to Figs 1–7 can be found in van Haren (2026), “Processed temperature data to: A note on semidiurnal motions in the deep Mediterranean”, Mendeley Data, V1, https://doi.org/10.17632/k2whybdm8c.1.

**Competing interests** The author has no competing interests.

## Appendix A Scheme of the experiment

An outlay of the large-ring mooring is given in Fig. A1. The height of the experiment is about twice the size of the horizontal extent (Fig. A1a). The large steel-tube ring has a diameter of nearly 70 m which functioned as a 140-kN anchor (Fig. A1b). The eighteen 12-m long and 0.6-m diameter steel tubes hold a grid of steel cables for rigidness. The cables are 9.5 m apart. At their intersections, 2.5-m diameter ‘small’ rings are mounted that each hold a 125-m long vertical mooring line with 65 T-sensors below a single 1.45-kN buoy. Of eight small rings, imaginary intersects are at the steel tubes, so that special off-set mounting is needed with three assist cables (details in van Haren et al. 2021).

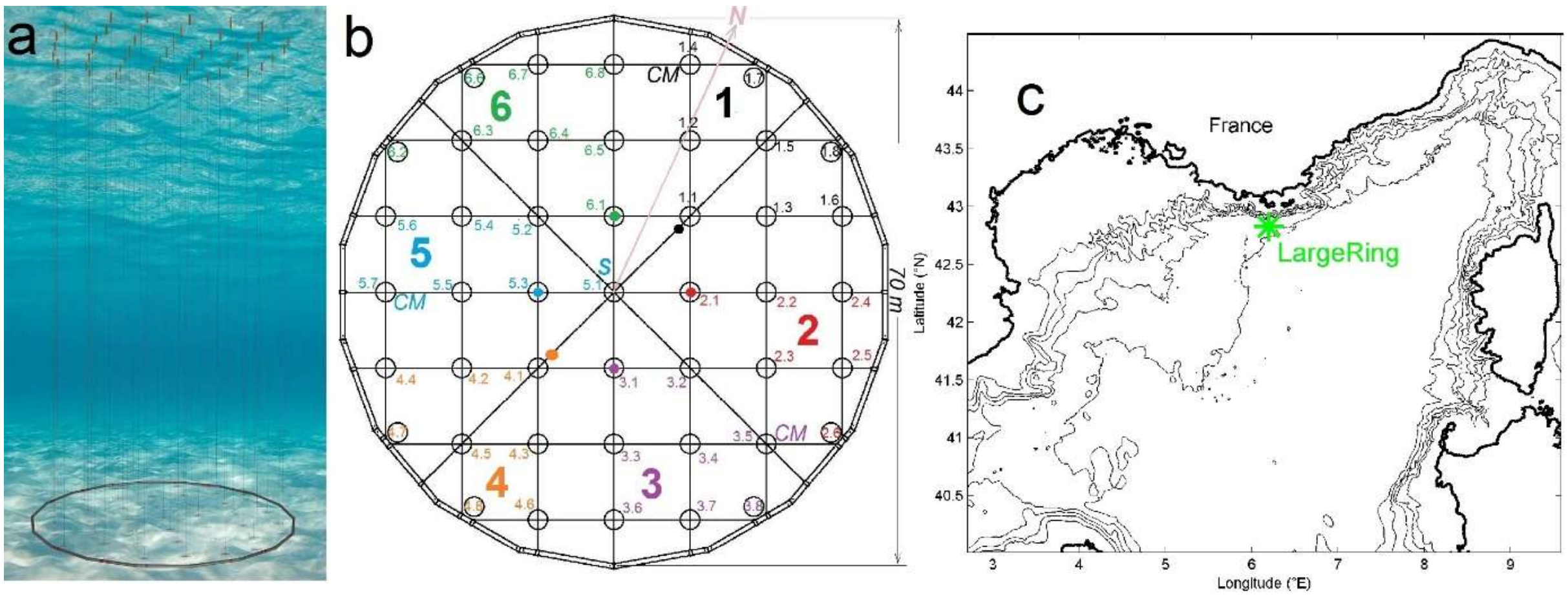

**Figure A1.** Scheme of the experiment. (a) Artist impression of the fully unfolded large-ring mooring underwater (courtesy J. van Bennekom). The sketch has an approximately correct vertical/horizontal aspect ratio. (b) Orientation and layout of the large-ring mooring viewed from above, with steel-cable grid and small rings numbered in six synchronisation groups with colour dots indicating group nodes and synchroniser ‘S’ at line 51. Here and elsewhere in the text, lines are indicated without period for short. Lines 14, 35 and 57 held a waterflow current meter ‘CM’ at the buoy. (c) Area map of the Northwestern Mediterranean Sea with the site of the large-ring mooring indicated by the green asterisk. Black contours indicate isobaths every 500 m.

The ‘large-ring mooring’ was deployed at the <1° flat and 2458-m deep seafloor of 42° 49.50′N, 006° 11.78′E, 10 km south of the steep continental slope in the Northwestern Mediterranean Sea, in October 2020. Upon landing at the seafloor following parachute-controlled ‘free’ fall, the orientation of the ring was directed to the NNW, pointing at 337° N (Fig. A1b). The continental slope roughly guides the Northern Current, a meandering waterflow marking the boundary between coastal and open-basin waters (e.g., Crepon et al. 1982).

The vertical mooring lines are named in six synchronisation groups, of maximum eight lines each. The single synchroniser is located at the small ring of central line 51 (Fig. A1b). Every half hour, the synchroniser sends a clock pulse to a group. The synchronisation sequence of six pulses is repeated every four hours.

**References**

Albérola C, Millot C, Font J (1995) On the seasonal and mesoscale variabilities of the Northern Current during the PRIMO-0 experiment in the western Mediterranean Sea. Oceanol Acta 18:163-192

Crepon M, Wald L, Monget JM (1982) Low frequency waves in the Ligurian Sea during December 1977. J Geophys Res 87:595-600

Ferron B, Bouruet Aubertot P, Cuypers Y, Schroeder K, Borghini M (2017) How important are diapycnal mixing and geothermal heating for the deep circulation of the Western Mediterranean? Geophys Res Lett 44:7845-7854

Gerkema T, Zimmerman JTF, Maas LRM, van Haren H (2008) Geophysical and astrophysical fluid dynamics beyond the traditional approximation. Rev Geophys 46:RG2004

IOC, SCOR, IAPSO (2010) The International Thermodynamic Equation of Seawater – 2010: Calculation and Use of Thermodynamic Properties, Intergovernmental Oceanographic Commission, Manuals and Guides No. 56, UNESCO, Paris, 196 pp

Kunze E (1985) Near-inertial wave propagation in geostrophic shear. J Phys Oceanogr 15:544-565

LeBlond PH, Mysak LA (1978) Waves in the Ocean. Elsevier, New York, 602 pp

Marshall J, Schott F (1999) Open-ocean convection: Observations, theory, and models. Rev Geophys 37:1-64

Sheremet VA (2004) Laboratory experiments with tilted convective plumes on a centrifuge: A finite angle between the buoyancy force and the axis of rotation. J Fluid Mech 506:217-244

Straneo F, Kawase M, Riser SC (2002) Idealized models of slantwise convection in a baroclinic flow. J Phys Oceanogr 32:558-572

Testor P, Gascard JC (2006) Post-convection spreading phase in the Northwestern Mediterranean Sea. Deep-Sea Res 53:869-893

van Haren H (2018) Philosophy and application of high-resolution temperature sensors for stratified waters. Sensors 18:3184

van Haren H (2022) Thermistor string corrections in data from very weakly stratified deep-ocean waters. Deep-Sea Res I 189:103870

van Haren H, Millot C (2003) Seasonality of internal gravity waves kinetic energy spectra in the Ligurian Basin. Oceanol Acta 26:635-644

van Haren H, Groenewegen R, Laan M, Koster B (2005) High sampling rate thermistor string observations at the slope of Great Meteor Seamount. Ocean Sci 1:17-28

van Haren H, Adrián-Martínez S, Al Samarai I, Albert A, André M, et al (2014) High-frequency internal wave motions at the ANTARES site in the deep Western Mediterranean. Ocean Dyn 64:507-517

van Haren H, Bakker R, Witte Y, Laan M, van Heerwaarden J (2021) Half a cubic hectometer mooring-array 3D-T of 3000 temperature sensors in the deep sea. J Atmos Ocean Technol 38:1585-1597

van Haren H, Adriani O, Albert A, Alhebsi AR, Alshalloudi S, et al (2026) Whipped and mixed warm clouds in the deep sea. Geophys Res Lett 53:e2025GL119998

Wunsch C, Ferrari R (2004) Vertical mixing, energy, and the general circulation of the oceans. Annu Rev Fluid Mech 36:281-314